\documentclass{emulateapj}
\usepackage{amssymb}
\usepackage{graphicx}
\usepackage{amsmath}
\usepackage{natbib}
\usepackage{epsfig}
\usepackage{epstopdf}

\usepackage{mathrsfs}
\usepackage{amsfonts}
\shorttitle{High z AGNs in C-COSMOS} \shortauthors{Civano et al.}

\def\xmm{{XMM-{\it Newton}}}
\def\chandra{{\it Chandra}}
\newcommand{\cgs}{ ${\rm erg~cm}^{-2}~{\rm s}^{-1}$} 
\newcommand{\lum}{\rm erg~s$^{-1}$}

\begin{document}

\title{The Population of High Redshift AGNs in the Chandra COSMOS survey}


\author{F. Civano\altaffilmark{1}, M. Brusa\altaffilmark{2}, A. Comastri\altaffilmark{3}, 
M. Elvis\altaffilmark{1}, M. Salvato\altaffilmark{4}, G. Zamorani\altaffilmark{3}, P. Capak\altaffilmark{5}, 
F. Fiore\altaffilmark{6}, R. Gilli\altaffilmark{3}, H. Hao\altaffilmark{1}, H. Ikeda\altaffilmark{7}, 
Y. Kakazu\altaffilmark{5},  J. S. Kartaltepe\altaffilmark{8},
D. Masters\altaffilmark{5,9}, T. Miyaji\altaffilmark{10}, M. Mignoli\altaffilmark{3}, S. Puccetti\altaffilmark{11},
F. Shankar\altaffilmark{12}, J. Silverman\altaffilmark{13}, 
C. Vignali\altaffilmark{14},  A. Zezas\altaffilmark{1,15}, A. M. Koekemoer\altaffilmark{16}} 

\altaffiltext{1}{Harvard Smithsonian Center for astrophysics, 60 Garden St., Cambridge, MA 02138, USA}
\altaffiltext{2}{Max Planck Institut f\"ur extraterrestrische Physik Giessenbach strasse 1, D--85748 Garching, Germany}
\altaffiltext{3}{INAF-Osservatorio Astronomico di Bologna, via Ranzani 1, I-40127 Bologna, Italy}
\altaffiltext{4}{Max-Planck-Institute for Plasma Physics, Boltzmannstrass 2, Garching D--85748, Germany} 
\altaffiltext{5}{California Institute of Technology, 1201 East California blvd, Pasadena, CA 91125, USA}
\altaffiltext{6}{INAF-Osservatorio Astronomico di Roma, via Frascati 33, Monteporzio-Catone (Roma), I-00040, Italy}
\altaffiltext{7}{Ehime University, 2-5 Bunkyo-cho, Matsuyama 790-8577, Japan}
\altaffiltext{8}{Institute for Astronomy, 2680 Woodlawn Dr., University of Hawaii, Honolulu, Hawaii, 96822}
\altaffiltext{9}{Department of Physics and Astronomy, University of California, 900 University Ave, Riverside, CA 92521, USA}
\altaffiltext{10}{Universidad Nacional Autonoma de Mexico-Ensenada, Km. 103 Carretera Tijuana-Ensenada, 22860 Ensenada, Mexico}
\altaffiltext{11}{ASI Science Data Center, via Galileo Galilei, 00044 Frascati, Italy} 
\altaffiltext{12}{Max-Planck-Institut fu\"r Astrophysik, Karl-Schwarzschild-Str. 1, D-85748 Garching, Germany}
\altaffiltext{13}{Institute for the Physics and Mathematics of the Universe (IPMU), University of Tokyo, Kashiwanoha 5-1-5, Kashiwa, Chiba 277-8568, Japan} 
\altaffiltext{14}{Dipartimento di Astronomia, Universit\'a di Bologna, via Ranzani 1, 40127, Bologna, Italy}
\altaffiltext{15}{Physics Department, University of Crete, P.O. Box 2208, GR-710 03, Heraklion, Crete, Greece}                                     
\altaffiltext{16}{Space Telescope Science Institute, 3700 San Martin Drive, Baltimore, MD 21218, USA}                                     



\begin{abstract}
We present the high redshift (3$<z<$5.3) 0.5-2 keV number counts and the 2-10 keV (rest frame) space density of 
X-ray selected AGNs detected in the \chandra\ COSMOS survey. The sample comprises 81 X-ray detected sources with available 
spectroscopic (31) and photometric (50) redshifts plus 20 sources 
with a formal $z_{phot}<3$ but with a broad photometric redshift probability distribution, such that $z_{phot}+1\sigma>3$.
81 sources are selected in the 0.5-2 keV band, 14 are selected in the 2-10 keV and 
6 in the 0.5-10 keV bands. We sample the high luminosity ($log~L_{(2-10 keV)}>$44.15~\lum) space density up to $z\sim$5 and a fainter luminosity range 
(43.5$<log~L_{(2-10 keV)}<$44.15~\lum) than previous studies, up to $z$=3.5. We weighted the contribution to the number 
counts and the space density of the sources with photometric redshift by using their probability of being at $z>$3. 
We find that the space density of high-luminosity AGNs declines exponentially at all the redshifts, confirming the trend 
observed for optically selected quasars. At lower luminosity, the measured space density is not conclusive, and a larger sample of 
faint sources is needed. Comparisons with optical luminosity functions 
and BH formation models are presented together with prospects for future surveys.
\end{abstract}


\keywords{Galaxies: active - evolution - surveys - X-rays: galaxies}



\section{Introduction}
To properly test models 
on the co-evolution of  black holes (BH) and galaxies (e.g., Granato et al. 2001, 2004; Croton et al. 2006; Hopkins et al. 2006; Menci et al. 2008), 
the accretion activity in the Universe has to be studied up to high redshifts and low luminosities. 
This requires large samples of Active Galactic Nuclei (AGNs) spanning wide ranges of properties.  

Although sizable samples have been collected in the optical and 
a decline in the quasar density between redshift $\sim$2.5 and 6 has been observed 
(Fan et al. 2001; Willott et al. 2003; Richard et al. 2006; Jiang et al. 2006), this evolution has been traced only by the most luminous 
($-$27.5$<$M$_{1450}<-$25.5) sources at redshift beyond 3. Only recently, Glikman et al. 
(2010, 2011) have given constraints on the faint end of the luminosity function in 
the optical band by using a sample of more than 40 faint (M$_{1450}<-$22) quasars in the 3.74 to 5.06 redshift range.
The presence of dust in high redshift sources (Jiang et al. 2006; Gallerani et al. 2010)
could, however, strongly affect their color selection and thus also the optical
luminosity function.

Thanks to the sensitivity reached by the \chandra\ and \xmm\ satellites, 
sizable samples ($\sim$40 sources in the largest one) of $z\sim$3-4 X-ray detected AGNs have been collected 
(Silverman et al. 2005, 2008; Brusa et al. 2009; Yencho et al. 2009; Ebrero et al. 2009; Aird et al. 2010; Fiore 2010). 
X-ray surveys are suitable for the selection of high-redshift low-luminosity AGNs because they are little affected by obscuration, unless 
Compton Thick absorption (Brandt \& Hasinger 2005). Indeed, only a few Compton Thick AGNs at z$>$3 have been recognized so far 
(Norman et al. 2002; Comastri et al. 2011; Gilli et al. 2011).
At the same time, X-ray surveys are limited in collecting large samples, 
either because wide area surveys have been not deep enough to detect $z>$4 AGNs or because deep surveys are too 
small and affected by cosmic variance. Only two $z>$5 spectroscopically confirmed quasars have been X-ray selected so far, one in the CDFN 
(z=5.19, Barger et al. 2005) and the other in the CLASXS survey (z=5.4, Steffen et al. 2004), both of them being unobscured type 1 quasars. 
 
The space density of luminous quasars in the XMM-COSMOS survey at 3$<z<$4 (Brusa et al. 2009) mimics the exponential decline observed 
in the optical. However, the higher redshifts and lower luminosities regimes are still unexplored. The faint X-ray luminosity range 
($\sim$10$^{43}$~\lum) is well sampled up to $z\sim$3 (see e.g., Hasinger 2008) but not many sources have been detected at this 
luminosity at higher redshifts. 

The goal of this paper is to compute the space density of high redshift 
AGNs in the 2-10 keV rest frame band that is little affected by obscuration (up to N$_H$=few $\times$10$^{23}$ cm$^{-2}$).
For this reason, the main selection has been performed in the 0.5-2 keV band, closely corresponding to the 2-10 keV rest frame band at $z>$3. 
In order to take into account for the presence of highly (up to N$_H \sim 10^{24}$ cm$^{-2}$) obscured AGNs, we also considered the sources 
with detection only in the observed 2-10 or 0.5-10 keV bands.
Taking advantage of the medium-depth, large-area \chandra\ survey of the COSMOS field (C-COSMOS, Elvis et al. 2009; Puccetti et al. 2009), 
we sample the 43.5$<log L_{(2-10 keV)}<$45~\lum\ luminosity range to study the number counts and 
space density of high (3$<z<$6.8) redshift AGNs. 

Throughout the paper we quote AB system magnitudes and we assume a cosmology with H$_{0}$ = 70~km~s$^{-1}$~Mpc$^{-1}$, 
$\Omega_M$ = 0.3 and $\Omega_{\Lambda}$= 0.7.

\section{High-redshift AGN sample}

 The \chandra-COSMOS survey (C-COSMOS; Elvis et al. 2009) covers the central 0.9 deg$^2$ of the COSMOS 
field to a depth of up 200 ksec in the inner 0.5 deg$^2$. The C-COSMOS X-ray source catalog comprises 1761 point-like X-ray sources 
detected down to a maximum likelihood threshold detml$=$10.8 at least in one band (0.5-2, 2-8 and 0.5-8 keV). 
This likelihood threshold corresponds to a probability of $\sim 5 \times 10^{-5}$ that a catalog source is instead a 
background fluctuation (Puccetti et al. 2009). Given the likelihood
threshold above, the flux limit reached in the survey is 5.7$\times$10$^{-16}$~erg~cm$^{-2}$~s$^{-1}$ in
the Full band (0.5--10~keV), 1.9$\times$10$^{-16}$~erg~cm$^{-2}$~s$^{-1}$ in the
Soft band (0.5--2~keV) and 7.3$\times$10$^{-16}$~erg~cm$^{-2}$~s$^{-1}$ in the
Hard band (2--10~keV). 

The high redshift AGN sample used in this work has been selected from the C-COSMOS X-ray catalog, 
combining the spectroscopic and photometric information available from the 
identification catalog of the 1761 X-ray C-COSMOS sources (Civano et al. 2011 in prep.).  
 
First, we selected all the sources with secure spectroscopic 
redshift from the identification catalog (870 sources). 
The brightest C-COSMOS sources, quite often associated with XMM-COSMOS sources, have been observed as compulsory targets by the zCOSMOS 
bright (VIMOS/VLT, Lilly et al. 2007, 2009) and the Magellan/IMACS (Trump et al. 2007, 2009) surveys limited to i$<22.5$. 
Recently, the C-COSMOS sources have been the primary targets of much deeper observations with the Keck/DEIMOS 
(to i$<$25\footnote{The Keck survey is the result of a multi-year observing campaign (PIs: 
Capak, Kartaltepe, Salvato, Sanders, Scoville).}) and VIMOS/VLT (zCOSMOS deep, to B$<$25) surveys. 
In this spectroscopic sample, 29 sources with soft band X-ray detection have redshift larger than 3, 
of which 6 are at $z>$4 and 2 at $z>$5. Only two sources with $z_{spec}>$3 do not have soft band detection, but only hard band detection.
The highest redshift source with an optical spectrum, at $z_{spec}$=5.3, 
is one of the 3 spectroscopically confirmed members of the high-$z$ proto-cluster discovered 
in the COSMOS field, and its optical spectrum is reported in Capak et al. (2011). This 
quasar is the only X-ray detected object in the proto-cluster.

In summary, the spectroscopic sample includes 31 sources. 

Second, given that only $\sim$50\% of the C-COSMOS sources have a spectroscopic 
redshift, the sources with photometric redshifts (Salvato et al. 2009, 2011 submitted), which are  
typically fainter (i$_{AB}$=21.5 and 23.8 are the mean optical magnitudes of the spectroscopic 
and photometric samples, respectively), need to be included, as these extend the sample 
to low luminosities and high redshifts. Given the large number of photometric bands 
(31, of which 12 are intermediate bands, suitable for the selection of emission lines), 
the spectral energy distribution of COSMOS sources can be used as a low-resolution spectrum. 
The COSMOS photometric redshifts for X-ray selected sources have an accuracy of 
$\sigma_{\Delta z/(1+z_{spec})}$=0.015 with a low number of outliers ($<$6\%), considering the sample as a whole. 
These numbers depend on the magnitude of the sources, but at z$>$2.5 a remarkably good accuracy of $\sigma_{\Delta z/(1+z_{spec})}$=0.011 is achieved.
The spectral energy distributions (SEDs) of the sources with photometric redshift larger than 3 have been visually 
inspected together with the photometric fitting and the probability distribution of all the possible solutions. 

This process adds 36 sources with photometric redshift larger than 3 and a soft band detection (7 are at $z>$4 
and 2 at $z>$5), plus 10 sources with hard and full band detection and 4 sources with a full band detection only. 
Adding the sources with photometric redshifts, we more than double the spectroscopic sample at each redshift (Table~1). The source with the highest 
photometric redshift is CID-2550 (soft band detected only), for which a photometric redshift of $z_{phot}$=6.8 has been computed 
(see Salvato et al. 2011, submitted). 
To account for the broad and multiply peaked shape of the photometric redshift probability distribution 
({\it P(z)}) and thus the possible contamination due to low redshift sources, we computed the 
fraction of the {\it P(z)} at $z_{phot}>$3 for each source with a photometric redshift only, 
to weight their contribution to the number counts and space density computation.   
73\%\ of these sources have a {\it P($z_{phot}>$3)} larger than 0.5.

From the C-COSMOS identification catalog we also 
selected those sources detected in the soft, hard or full band having a broad {\it P(z)}, 
such that $z_{phot}+ \sigma_{zphot}>$3 and $z_{phot}<$3. 
Their SEDs have been visually 
inspected to check for possible problems, such as contaminated photometry due to nearby objects. A total of 20 sources 
(16 detected in the soft band, 2 in the hard and full band and 2 in the full band only) were added to the main sample. 
We adopted $z=z_{phot}+ \sigma_{zphot}$, and weighted their contribution to the number counts and to the space density by 
the fraction of the {\it P(z)} at redshift $>3$. Only 15\%\ of these sources have a {\it P($z_{phot}>$3)} larger than 0.5.

Of the 81 soft band detected high-z sources, 32 are also XMM-COSMOS sources (Brusa et al. 2010). 
Half were included in the Brusa et al. (2009) paper on $z>$3 sources,  while 5 sources were below the XMM-COSMOS soft band 
threshold adopted for that study (10$^{-15}$ \cgs). For the remaining 11 sources either a new 
version of the photometric redshift catalog (Salvato et al. 2011, submitted), which employs a new and improved SED 
library and the addition of deep H band photometry, 
suggests a higher $z$ solution (7 sources), or a newly available spectroscopic redshift (lower than $z<3$) superseded 
the previously available photometric one (4 sources). Brusa et al. (2009) also 
did not include the hard band detected sources.
The new photometric redshifts (Salvato et al. 2011, submitted) have been tuned using the C-COSMOS sources which, being fainter 
than the XMM-COSMOS ones, are typically more galaxy-dominated and thus in few cases a different photometric redshift has been 
proposed with respect to Brusa et al. (2007, 2009). 

In summary, the total $z>$3 C-COSMOS sample includes 81 sources with $z>$3 (spectroscopic or photometric) plus 20 sources candidate to be at $z>$3 
from their broad {\it P(z)}. The effective size of the sample, computed by summing the fraction of the 
{\it P(z)} at $z>$3 for the 101 sources ({\it P(z)}=1 for the spectroscopic redshifts), is of 73 sources. 
In the total sample, 81 sources are soft band selected, 14 are hard band selected 
and 6 are full band selected (Table~1). The X-ray fluxes together with the redshifts of the sources are listed in Table~3 (see published version).

Therefore, the C-COSMOS high-redshift sample increases by a factor of 2 the Brusa et al. (2009) sample. 
This is the largest available sample of $z>3$ X-ray selected AGNs in a contiguous sky area, even when its effective size 
is considered. 
For comparison, in the 2Ms CDFS optical 
identification catalog (Luo et al. 2010, see also Silverman et al. 2010) there are 45 sources at $z>3$ 
(6 with spectroscopic and 39 photometric redshifts), of which 3 are at $z>4$, though none of these are spectroscopically identified. 
Considering the 4Ms CDFS catalog (Xue et al. 2011), 20 sources are added to the $z>3$ catalog, though with only photometric redshifts.
Fiore et al. (2011, submitted) will report the combination of the results obtained in all the above X-ray surveys.

\subsection{Optically Unidentified Sources}

There are also 18 C-COSMOS sources without a counterpart in the 
optical bands, but with a K-band and IRAC (10), only IRAC (6) or no detection (2). 
The optical images of these sources have been visually inspected to verify that their absence 
in the main optical catalog is not due to a source detection problem, and no optical emission has been found at the X-ray positions.
Given the small number of bands in which these objects are detected, no photometric redshift is available for them. 
In X-ray selected samples, non--detection in the optical band has been often assumed to be a proxy for 
high redshift (e.g. Koekemoer et al. 2004), or for high obscuration, or a combination of both.  

Four of the 18 sources have no detection in the soft band suggesting high obscuration, possibly  
combined with high redshift. 
We considered the 14 soft band detected sources to be at $z>$3 on the basis of their optical 
non-detection and we included them in the derivation of the upper boundary of the logN--logS curve (Section 3). 
Their contribution to the space density will be discussed in Section 4.

\begin{figure}
\centering
\includegraphics[width=0.49\textwidth]{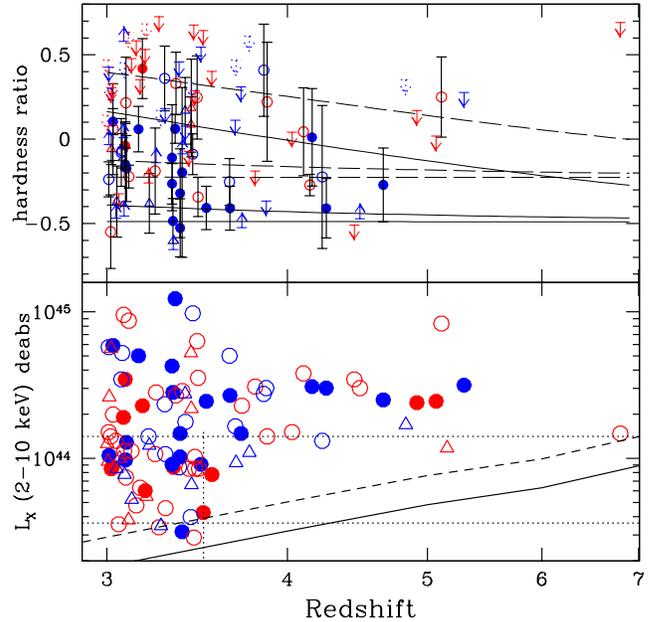}
\caption{{\it Top:} Hardness ratio versus redshift. Blue= type 1. Red = Not type 1 (see \S\ 2.3 for the definition). Filled = spectroscopic redshift. 
Open = photometric redshift. Sources with no hard band or soft band detection are reported as arrows. 
Three curves of constant N$_H$ (10$^{20}$, 5$\times$10$^{22}$ cm$^{-2}$ and 5$\times$10$^{23}$ cm$^{-2}$) 
are reported as dashed lines for $\Gamma$=1.4 and as solid lines for $\Gamma$=2. The sources with $z_{phot}+ \sigma_{zphot}>$3 
have been plotted as triangles 
or as dashed arrows if they do not have hard band or soft band detection. {\it Bottom:} The luminosity (computed with $\Gamma$=1.4) redshift plane
for the objects in our sample. The continuous line represents the de-absorbed 2--10 keV luminosity limit of the survey 
computed from the 0.5--2 keV limiting flux. The dashed line corresponds to the flux limit we imposed for the 
computation of the number counts and space density. Symbols are as in the top panel. The dotted 
black lines represent the luminosity limits used in the space density computation. }
\end{figure}

\subsection{X-ray properties}
\label{xraypro}
To allow a useful comparison of the space density with model predictions, 
obscuration has to be taken into account. The sources in our sample have a low number of 
detected counts (28 is the median value in the 0.5-7 keV band). 
In this count regime, spectral fit results are not stable, in particular when more than one free parameter is added to the fitting. Even if 
the fit converges the uncertainties on the parameters are large.  
For these reasons, we used the hardness ratio, defined as HR=(H-S)/(H+S) (where H and S are the counts in 
the 2-7 keV and 0.5-2 keV bands respectively), to give a rough estimate of the obscuration affecting the sources 
and derive the intrinsic luminosity. 
The upper or lower limits on the HR for the sources with detection only in one band (14 only hard and 37 only soft) 
have been computed by 
converting the 3~$\sigma$ flux upper limit, in the band in which these sources are not detected, into counts.  

To estimate the column density, curves of constant $N_H$ as a function of redshift have been derived for 
two spectral slope values, $\Gamma$=1.4 and $\Gamma$=2. The flat spectral slope has been chosen to be consistent with the assumptions 
adopted in producing the original X-ray catalog (Puccetti et al. 2009). The steeper value is more representative of the intrinsic 
value if the spectrum is not affected by obscuration (Nandra \& Pounds 1994). 

In Figure 1, top panel, the curves of N$_H$=10$^{20}$, 5$\times$10$^{22}$ cm$^{-2}$ and 5$\times$10$^{23}$ cm$^{-2}$ are reported 
for $\Gamma$=1.4 (dashed lines) and $\Gamma$=2 (solid lines). Though there is a group of sources with HR 
indicative of significant obscuration, the low number of counts, the large HR 
error and the similarity in this redshift range of the curves with widely different N$_H$ values for the same spectral slope, 
do not allow an accurate estimate of the column density for each source to be made. We also note that using two different intrinsic 
slopes implies quite different obscuration. 

We thus classified as absorbed, those sources whose $HR-1\sigma$  are above 
the value corresponding to N$_H$=5$\times$10$^{22}$ cm$^{-2}$ at their own redshift for each spectral index. 
This criterion gives 9  X-ray obscured sources (detected in both soft and hard bands) for $\Gamma$=1.4, 
or 32 sources for $\Gamma$=2. Hereafter, we will use $\Gamma$=2, which gives a better estimate of the sources intrinsic column density.
The luminosity of the candidate obscured sources have been corrected by using a correction factor derived from XSPEC (Arnaud 1996) 
by simulating an absorbed power-law with the N$_H$ computed above and $\Gamma$=2.   

The 37 soft band sources with no detection in the hard band (reported as downward arrows in Fig. 1, top panel) have a very high upper limit on the HR, 
due to the conservative flux upper limit computed by Puccetti et al. (2009). Their non-detection in the hard band is not necessarily 
due to obscuration but 
to the fact that these sources are very faint in the soft band and thus their hard band flux is below the flux limit of the survey. 

For the 81 soft-band detected objects, the 2-10 keV rest-frame luminosity was computed 
by using the 0.5-2 keV flux, which at $z>$3 corresponds to the hard ($>$2 keV) rest-frame emission, 
assuming $\Gamma$=2. If needed, the absorption correction has been applied.

The 2-10 keV rest frame luminosity for the 14 z$>$3 hard-band only detected sources (upward pointing arrows in Fig. 1, top panel) has been computed 
by converting the observed 2-10 keV flux into rest frame luminosity, using $\Gamma$=2.
Although the non detection in the soft band is suggestive of large column density ($>$5$\times$10$^{23}$ cm$^{-2}$), 
from the HR analysis only 4 of the HR lower limits lie above the N$_H$=5$\times$10$^{22}$ cm$^{-2}$ curve 
(for either choice of spectral slope assumptions), in the region of obscured sources. 

For the 4 full band only detected sources, the 2-10 keV rest frame luminosity has been computed by converting the full band 
flux and assuming the same spectral slopes. 

 The absorption-corrected hard X-ray luminosity (computed with $\Gamma$=2) versus redshift plane is reported in Figure~1 (bottom panel) together 
with the flux limit of the C-COSMOS survey (solid line).

\subsection{Optical properties}

The $z>$3 C-COSMOS spectroscopic sub-sample (31 sources) includes 19 broad line AGNs (type 1, FWHM$>$2000 km/s) and 12 AGNs with narrow lines 
only (not type 1). 
The brightest (i$_{AB} \sim$22 - 23) sources of the spectroscopic sample for which 
spectroscopic identification is available are type 1 AGNs. At fainter optical magnitudes (i$_{AB}>$23), we find equal number of 
type 1 and not type 1 AGNs.

From the photometric fitting, an SED type can be derived for each source without spectroscopic classification. About half of the total sample is
best fitted with an unobscured quasar template (type 1-like), and half with an obscured
quasar template (not type 1-like, see Salvato et al. 2009 for more details on the SED templates). 

The spectral types (solid circles) along with the SED fitting type (open circles) are reported in Fig.~1 
(type 1 in blue and not type 1 in red).

The optical spectra of X-ray selected AGN show a broader range of properties 
than optical selected samples. 
Even though a detailed analysis of the spectral properties (emission and absorption line intensity, continuum slope, extinction) 
will be the subject of a following paper, the variety of striking features seen in the spectra are briefly reported below. 

The sources classified as type 1 AGNs have spectra with broad (FWHM$\sim$5000 km/s) lines of either Ly$\alpha$ (1216 \AA) and CIV (1549 \AA), 
or CIV and [CIII] (1909 \AA). 

\begin{figure}
\centering
\includegraphics[width=0.49\textwidth]{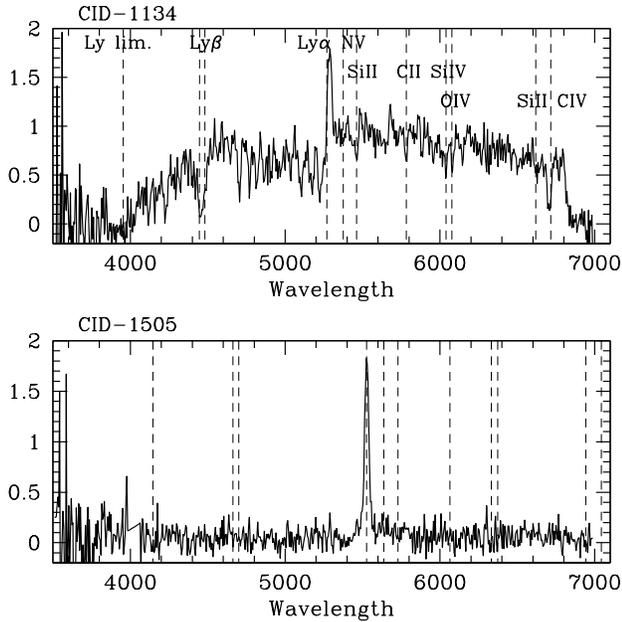}
\caption{Optical spectra of source CID-1134 (top, z=3.335) and CID-1505 (bottom, z=3.546) from the VLT deep survey of the COSMOS 
field (zCOSMOS deep, Lilly et al. 2009). The spectra are plotted in arbitrary flux normalization. The dashed lines mark the 
same lines in both spectra (Ly limit, Ly$\beta$, OVI, Ly$\alpha$, NV, SiII, CII, SiIV and OIV, SiII and CIV).}
\label{spectra}
\end{figure}

Among the narrow line sources, three sources (Fig. \ref{spectra}, top panel) show spectra typical of normal star-forming galaxies with 
a narrow Ly$\alpha$ emission and stellar absorption lines (Ly$\beta$, SiII at 1260\AA, CII at 1334 \AA, SiIV at 1393 \AA, OIV at 1402 \AA\ 
and CIV at 1549 \AA; Shapley et al. 2003). Interestingly,
there is no hint of CIV line in emission, a typical signature for
nuclear activity, in these objects. No spectra of this kind were found in the brighter XMM-COSMOS sample of z$>3$ AGNs (Brusa et al. 2009), while a 
few similar spectra have been found in the same redshift range in the ECDFS survey (Silverman et al. 2010), though with a lower 
signal-to-noise ratio. In particular, the source shown in Fig. \ref{spectra} (top panel) is not detected in the soft band but only in 
hard band, suggesting the presence of high obscuration, consistent with the absence of nuclear emission lines.   
 
Four sources show strong narrow (FWHM$\sim$1000 km/s) 
emission lines (mostly Ly$\alpha$) over a faint, almost zero, continuum as shown in Fig. \ref{spectra} (bottom panel). 
The strength of their Ly$\alpha$  (EW$_{rest} \sim$250 \AA) is comparable with the strongest lines found in Ly$\alpha$ emitter samples 
(see e.g., Murayama et al. 2007; Cassata et al. 2010 and references therein).  One of the sources, in this class, is the
 highest redshift narrow line source ($z_{spec}$=5.07; Ikeda et al. submitted) in the sample. 
Without X-ray data, objects like those shown in Figure~2 would have been easily missed by optically-based AGN surveys. 
The optical spectra themselves would have not allowed their classification as AGN.

Broad absorption lines are visible in the spectra of a couple of sources, typically in the CIV doublet at 1549 \AA. 
These indicate the presence of outflows, which can provide an important means of carrying material and energy out of the central region, 
and could thus be important for the studies of the feedback and for the enrichment of the IGM at high redshift. 

\begin{table}
\footnotesize
\centering
\caption{\small Summary of the z$>$3 sample.}

\begin{tabular}{l c c c | c c c |  c c c |c c c}
\hline \hline
	&\multicolumn{3}{c}{Total}& \multicolumn{3}{c}{Spec.} & \multicolumn{3}{c}{Phot.}& \multicolumn{3}{c}{Phot.$+$1$\sigma>$3}	\\
	&S   &H & F & S &H & F &S &H & F &S &H & F	\\
z$>$3	& 81 &14 &6  &29 &2  &0  &36 &10 &4 &16&2 &2	\\
z$>$4	& 14 &1 &1  &6	&0  &0  &7  &1 &0  &1 &0 &1	\\	
z$>$5	& 4  &0 &1  &2	&0  &0  &2  &0 &0  &0 &0 &1	\\	

\hline \hline
	&\multicolumn{3}{c}{Total}& \multicolumn{3}{c}{Spec.} & \multicolumn{3}{c}{Phot.}& \multicolumn{3}{c}{Phot.$+$1$\sigma>$3}	\\
	&S   &H & F & S &H & F &S &H & F&S &H & F\\
z$>$3	& 73 &8 &3  &28 &1  &0  &30 &6 &2 &15&1 &1	 \\
z$>$4	& 13 &0 &0  &6	&0  &0  &6  &0 &0 &1 &0 &0	 \\	 
z$>$5	& 3  &0 &0  &2	&0  &0  &1  &0 &0 &0 &0 &0	 \\	 

\hline \hline
\end{tabular}

\vspace{0.3cm}
{\footnotesize {\it Top:} Summary of the sources belonging to the high redshift sample with secure spectroscopic,  
photometric redshift, and with $z_{phot}+ \sigma_{zphot}>3$. {\it Bottom:} Number of sources included in the space density computation, after applying 
a cut in flux limit.}
\end{table}

\section{Number counts of $z>3$ AGNs}

We derived the soft band logN--logS of the $z>3$ and $z>4$ C-COSMOS samples 
by folding the observed flux distribution through the sky coverage area versus flux curve 
of the C-COSMOS survey (Puccetti et al. 2009). 

 In order to minimize the error associated with the most uncertain part of the sensitivity 
curve, we truncate the sample at the flux corresponding to 10\% of the total area (dashed line in Fig.~1).
All the sources with a 0.5--2 keV flux above 3$\times$10$^{-16}$ \cgs\ have been considered (73 objects out of 81; 
column 1 of the lower Table 1).

The flux limit applied to the sample is consistent with the signal-to-noise ratio thresholds chosen by Puccetti et al. (2009), 
on the basis of 
extensive simulations, to avoid the Eddington bias in the computation of the number counts of the entire C-COSMOS sample. 
Thus, by applying a flux limit cut, we also reduce the Eddington bias affecting our sample.

\begin{figure}
\centering
\includegraphics[width=0.49\textwidth]{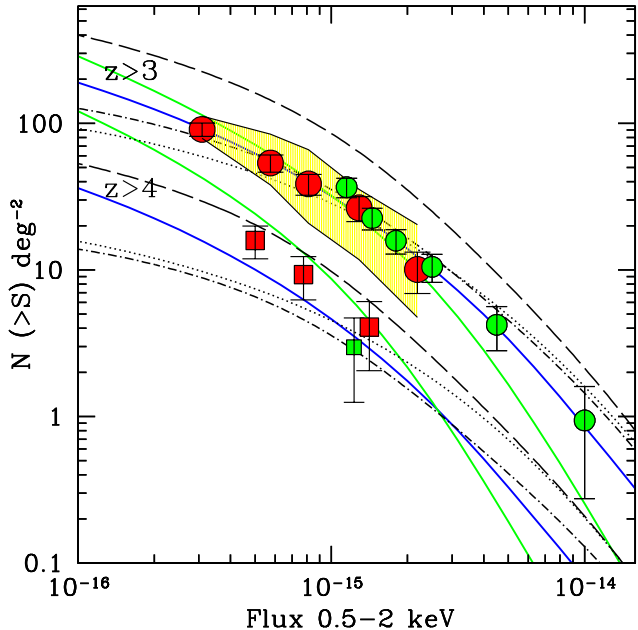}
\caption{The binned logN--logS relation (with associated
Poissonian errors) of the $z>3$ (red circles) and $z>4$ (red squares) QSOs population. 
The yellow shaded area represents the maximum and minimum number counts under the assumptions described in 
Section 3. The blue and green curves correspond to the prediction based on the Gilli et al. (2007)
and the Aird et al. (2010, steeper curves) models, respectively. The green symbol represents the number counts for $z>3$ (circles) and $z>4$ (squares) 
QSO estimated as in Brusa et al. (2009) for the XMM-COSMOS survey by using the most 
recent identification catalog (Brusa et al. 2010). The black dashed, dotted and dot-dashed lines represent different
predictions of a basic model of quasar activation by major mergers of dark matter haloes (see description in \S\ 3). }
\label{lnls}
\end{figure}

The binned logN--logS relations for two redshift ranges ($z>3$ and $z>4$) are plotted in Figure~\ref{lnls} (red points, with associated 
Poissonian errors), together with the XMM-COSMOS number counts (Brusa
et al. 2009, green points), estimated by using the most 
recent XMM-COSMOS identification catalog (Brusa et al. 2010). 

 The yellow shaded area represents an estimate of the maximum and minimum number
counts relation at $z >$ 3 obtained by considering two different effects. 
First, we computed the 1~$\sigma$ uncertainty in the sky coverage area for each source using the sky coverage as a function of flux of the 
C-COSMOS survey (see Figure 3 in Elvis et al. 2009) and the 1~$\sigma$ 
uncertainty in the flux (computed as in Puccetti et al 2009). 
This uncertainty is particularly important for sources with faint fluxes, where the error on the flux is large and 
the sky coverage curve is steep.
Second, we considered the 14 sources with no-optical detection (seen in the soft band).

To compute the upper boundary of the shaded area, we included all the sources in the main sample (73 objects) plus the sources with 
no optical detection at their flux$+$1$\sigma$ error.
For the lower boundary, we used the flux$-$1$\sigma$ error for all the sources in the main sample.

 The blue and green solid curves ($z>$3 upper and z$>4$ lower) in Figure~\ref{lnls} correspond to the predictions
of two different phenomenological models. 
The first (flatter solid blue curve) is the XRB synthesis model of Gilli et al. (2007\footnote{This model has been computed by using the 
POrtable Multi Purpose Application (POMPA) for AGN counts available 
at the url http://www.bo.astro.it/$\sim$gilli/counts.html.}), based on the 
X-ray luminosity function observed at low redshift (e.g., Hasinger, Miyaji \& Schmidt 2005), 
parametrized with a luminosity dependent 
density evolution (LDDE), and a high redshift exponential decline with the
same functional form adopted by Schmidt et al. (1995; 
$\Phi(z) = \Phi(z_0) \times 10^{-0.43(z-z_0)}$ and $z_0$=2.7) to fit the optical luminosity
function between z$\sim$2.5 and z$\sim$6 (Fan et al. 2001), corresponding to one e--folding per unit redshift. 
The second one (steeper green solid) is the luminosity and density evolution model (LADE; Aird et al. 2010) which fits 
the hard X-ray luminosity function derived by Aird et al. (2010) using the 2Ms \chandra\ Deep Fields 
and the AEGIS-X (200 ks) survey to probe the faint end (log~L$_X < 43$~\lum) and the high-z ($z\sim$3) range.

At $z >$ 3, the C-COSMOS points confirm and tighten the agreement with the model predictions, 
previously found by Brusa et al. (2009), extending this agreement to fainter fluxes. 
At redshift $z>4$, where the XMM-COSMOS sample had only 4 sources, the 
C-COSMOS sample is 3-4 times larger, making it possible to compare the slope of the counts with models, which agree. 

While at $z >$3 the two model predictions are very close, at $z>$ 4, where the models have different slopes, 
the errors on the data do not allow a firm preference of one of the two models.

While the above models are purely phenomenological, it is also possible to compare with 
physical models of quasar evolution.
In Figure~3, we compare the number counts with the prediction of a basic model of quasar activation 
by major mergers of dark matter haloes.  Briefly, the model consists of a DM halo merger rate compatible with
cosmological simulations and a quasar light curve that depends on the mass of the host and 
which describes the evolution of individual quasars (e.g., Wyithe \& Loeb 2003; Lapi
et al. 2006; Shen 2009; Shankar 2009, 2010; Shankar et al. 2010). 
The initial mass of the BH at triggering is assumed to be a fixed fraction
of its mass at the peak of activity. BH growth is regulated by a condition between
the peak luminosity and the mass of the host halo at the triggering time. 
In Figure \ref{lnls}, we report the
number counts obtained for each redshift range for a 
model with the same parameters as in Shen (2009; black dashed line) with 
minimum halo mass of $\sim$4$\times$10$^{11}$ M$_{\odot}$ (black dashed line).
The dot-dashed line refers to a model with larger minimum host halo mass 
($\sim$3$\times$10$^{12}$ M$_{\odot}$, dot-dashed line) presented in 
a preliminary work by Shankar (2010). 
In both models, a prolonged quasar light curve (evolving as t$-\alpha$ with $\alpha$=2.5-3), 
characterized by a long sub-Eddington post-peak phase, has been assumed.
The curve from a model characterized by a
lower minimum host halo masses (10$^{12}$ M$_{\odot}$) and negligible
post-peak phase (dotted line, Shankar 2010 for details) is reported too. 
The discussion of this comparison is reported in \S\ 5.

\begin{figure*}
\includegraphics[width=0.45\textwidth]{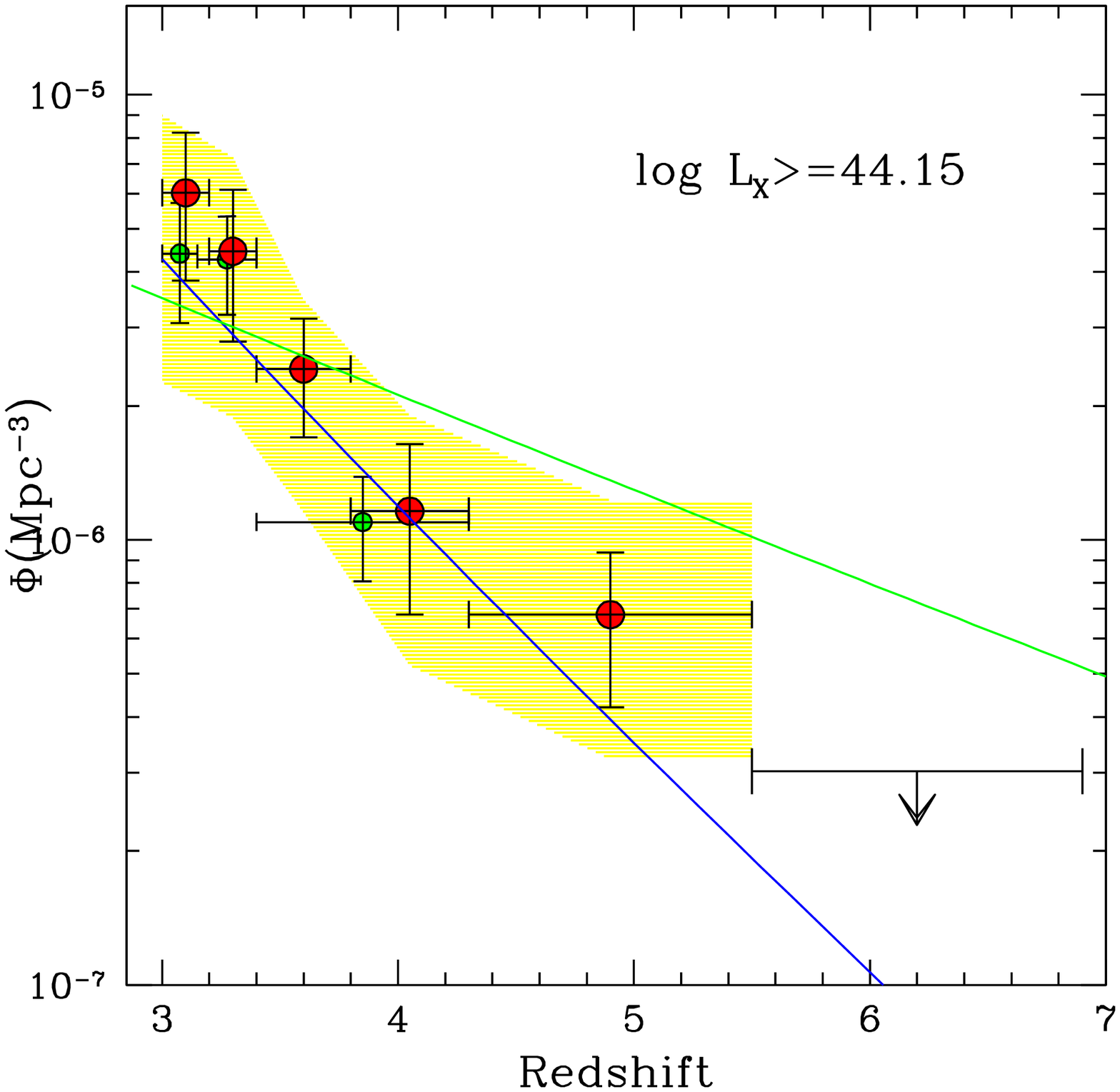}
\includegraphics[width=0.45\textwidth]{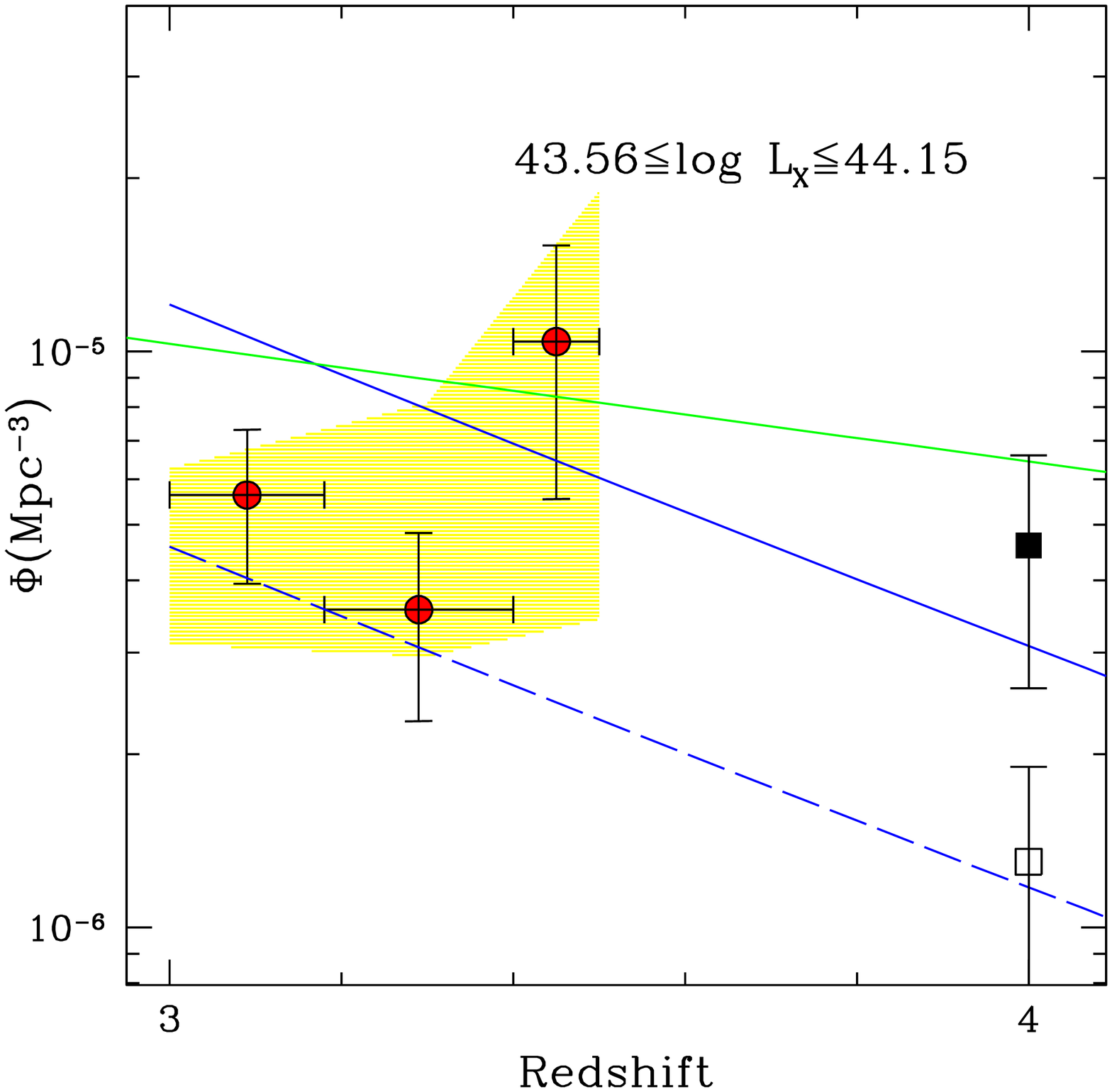}
\caption{ {\it Left:} The comoving space density in 6 different  
redshift bins at bright 2-10 keV X-ray luminosity, computed taking into account the effect of obscuration.
The blue curve corresponds to the X--ray selected AGN space density computed for
the same luminosity limit from the Gilli et al. (2007) model. 
The green curve corresponds to the space density derived from the LADE model of Aird et al. (2010). 
The yellow shaded area represents the maximum and minimum space density under the assumptions described in 
Section 3 and 4. The green symbols correspond to the data of Brusa et al. (2009) at luminosity $>$10$^{44.2}$ \lum.
{\it Right:} The comoving space density in 3 redshift bins for the low luminosity interval. The black open and solid
squares correspond to the luminosity function at $z\sim$4 of Ikeda et al. (2011) and Glikman et al. (2010, 2011), respectively, integrated in 
the M$_{1450}$=--21.8 -- --23.5 luminosity range. The dashed blue line corresponds to the Gilli et al. (2007) model if only unobscured sources 
are considered.}
\label{space}
\end{figure*}

\section{2-10 keV Comoving Space Density}

In order to avoid the Eddington bias at the faintest fluxes, we also applied a cut at the flux corresponding to 10\% of the total 
area to the hard and full band flux.
The fluxes associated with this area are 2$\times$10$^{-15}$ \cgs\ in 
the hard band and 1.2$\times$10$^{-15}$ \cgs\ in the full band. 
The number of sources used in the derivation of the space density is reported in Table~1 (bottom) 
in each band.

Including soft, hard and full band detected sources allows us to compute a space density which 
takes into account both unobscured, emitting more at softer energies, and obscured sources, emitting at more at harder energies, 
as shown in Section \ref{xraypro}, without having to introduce any further correction or assumption. 

The comoving space densities were computed with the 
1$/V_{max}$ method (Schmidt 1968), which takes into account the fact that more luminous objects 
are detectable over a larger volume.  The statistical uncertainty has been computed following Marshall et al. (1983).  
We also used the method proposed by Avni \& Bachall (1980) to take into account for the fact that each object could have 
been found in any point of the survey, thus at a different X-ray depth.
The maximum available volume, over which each source can be detected, has been computed 
by using the formula 

\begin{displaymath}
V_{max}=\int_{z_{min}}^{z_{max}} \Omega(f(L_X,z,N_{H}))\frac{dV}{dz}dz
\end{displaymath}

where $\Omega(f(L_X,z,N_{H}))$ is the sky coverage at the flux $f(L_X,z)$
corresponding to a source with absorption column density N$_H$ 
and unabsorbed luminosity L$_X$, and z$_{max}$ is the maximum redshift
at which the source can be observed at the flux limit of the
survey. If $z_{max}>z_{up,bin}$, then $z_{max}=z_{up,bin}$, where $z_{up,bin}$ is the upper boundary of the bin. 
More specifically, for unabsorbed sources we adopted
the observed rest-frame 2-10 keV luminosity,
while for obscured ones the unabsorbed luminosity was derived
assuming the best fit column density as obtained from the HR (as explained in Section \ref{xraypro}). 
We computed the space density using the luminosities derived with $\Gamma$=2. 
The contribution of sources with photometric redshift to the space density is weighted 
for the fraction of {\it P(z)} at $z>3$ (see $\S$2).

The comoving space density is shown in Figure~\ref{space}. 
In order to match the flux limit imposed above (see Fig.~1, dashed line) 
and to have a complete sample over a given redshift range, we divided the sample in two luminosity intervals.
At high luminosity (left panel), we computed the space density in 
6 redshift bins ($z=3-6.2$) at $log~L_X>$44.15 \lum, while for the low luminosity sample (right panel)
a cut at $log~L_X = 43.56$~\lum\ and $z$=3.5 (dotted lines in Fig.~1)
has been imposed.  
In Table~2, the number of sources in each redshift bin for the two luminosity ranges are
reported.

We also estimated space density upper and lower boundaries by taking into account the X-ray flux errors. 
If a source has been excluded from the main sample because its flux is lower 
than the flux limit applied, the same source could be included in the upper 
boundary sample if its flux$+$1$\sigma$ error exceeds the flux limit. 
Likewise, if a source has been included in the main sample because its flux is higher 
than the flux limit applied, the same source could be excluded by the lower boundary sample 
if its flux$-$1$\sigma$ error is lower than the flux limit. 
As for example, the source with photometric redshift z=6.8 (CID-2550) has a soft band flux below the chosen flux limit 
but the flux$+$1$\sigma$ error exceeds this limit so it is included in the upper boundary and being the only source in that 
redshift bin it has been plotted as an upper limit (at 3$\sigma$).

The yellow shaded area includes the above uncertainties affecting the computation of the space density, i.e., 
the flux errors and thus errors on the maximum volume associated to each source. 
In columns 3 and 4 of Table~2, the number of sources when the flux errors are considered are reported. 

As explained in Section 2.3, the 18 sources with no optical band detection 
have not been included in the space density boundaries. However, we computed the space density in 
the assumption that all the 18 sources were at the redshift corresponding to the first bin, then to the second bin and so on. 
The space density values computed in this case, in the first three bins, are within the yellow shaded area. If all the no-optically identified 
sources are at $z=$4.05 or at $z=$4.9, respectively, the number of sources per bin will be 4 times higher than the 
value reported in Table~2, and the same would be for the space density. However, this last unlikely option is the 
most extreme and we have not included this correction in the derivation of the upper boundary in these two high redshift bins.

The space density in the two luminosity ranges is compared with the 
predictions, at the same luminosity threshold, 
from the same Gilli et al. (2007) model used for the logN--logS (blue solid line), including in the model all the sources 
up to a column density of 10$^{24}$ cm$^{-2}$. We also compare with the LADE model (Aird et al. 2010; green solid line). 

At z$\sim$3, as already found for the number counts, the Gilli et al. (2007) and Aird et al. (2010) declining
space density provides a good representation of the observed data.
At $z>$4, the LADE model over-predicts the observed data in the 
high luminosity range, but, when the uncertainties are taken into account (shaded yellow area), the data lies in between 
the two models. 

No X-ray space density for these redshifts has been reported previously at low luminosities (43.56$<$log~L$_X <44.15$~\lum; 
Fig. \ref{space}, right). The size of the sample (32 sources), however, does not allow us to discriminate between models. 


 
In order to compare our data with other recent observations, we derived the space density at similar luminosity 
from the optical (at 1450 \AA) luminosity functions of broad line quasars at z$\sim$4 of Glikman 
et al. (2010, 2011) and Ikeda et al. (2011). For both authors, we used their two power-law luminosity function, 
since we are probing the region around M$_{\star}$ where strong curvature is present and both slopes contribute to the shape. 
Assuming a relation between the X-ray luminosity at 2 keV and the luminosity at 1500\AA\ 
($\alpha_{ox}=1.929-0.119 log L_{1500 \AA}$; Young et al. 2010), we converted the 43.56$<$log~L$_X <44.15$~\lum\ range 
(in the rest-frame hard band) into absolute magnitude at 1500\AA. We then integrated their luminosity functions 
between M$_{1450}$ = --21.8 and --23.5 to obtain the corresponding space density at $z\sim$4. The two values are reported 
in Figure~4 (right; Glikman et al. 2011 as a filled square, Ikeda et al. 2011 as an open square). 

The space densities derived from the two studies differ by a factor of three (as reported in Ikeda et al. 2011), and the reason 
for this difference is not well understood. It should be noted that both estimates are based on rather small samples 
(40 sources in Glikman et al. 2011 and 8 sources in Ikeda et al. 2011) and thus large errors should be taken into account. 
Though, the Glikman et al. (2011) result is consistent with the predictions of the Gilli et al. (2007) model when the obscured sources 
are taken into account (blue solid line) and the Aird et al. (2010) model, while the space density derived from the Ikeda et al. (2011) 
is in agreement with the prediction when only unobscured sources are considered (dashed line; Gilli et al. 2007), 
it is not clear how to explain the difference given that both samples include only broad line quasars. However, if we consider 
the optical type for our sources in each bin, we do not find a good agreement with the models too: in the first bin the majority of sources 
are not type 1 (see Table 2 column 2), but the space density value is closer to the prediction when only unobscured sources 
are considered (dashed line). 

 \begin{table}
\footnotesize
\caption{\small Summary of the sources in each luminosity and redshift bin. }\centering
\begin{tabular}{l c c c c  }
\hline \hline
z bin	&\multicolumn{4}{c}{Number of sources} \\ 
	&\multicolumn{4}{c}{log~L$_{(2-10keV)}>$44.15~\lum}\\
	\hline	 
	&N& N(Type)$^a$ &	N$_{lower}^{aa}$& N$_{upper}^{aa}$		\\
3.1 	&15& 5--10   &10    &21 				\\
3.3 	&10& 7--3    &6     &13				\\
3.6 	&12& 6--6   &8     &13					\\
4.05 	&6 & 3--3   &5     &8					\\ 
4.9	&7 & 3--4   &5     &9					\\ 
6.2     &0 & 0   &0	    &1					\\
\hline \hline 
z bin	&\multicolumn{4}{c}{Number of sources}\\
	& \multicolumn{4}{c}{43.56$<$log~L$_{(2-10keV)}<$44.15~\lum}\\
\hline 
       &N& N(Type)$^a$ & N$_{lower}^{aa}$& N$_{upper}^{aa}$ \\
3.09   &15& 5--10&   12 &13\\
3.29   &8&  5--3&  9  &11 \\
3.45   &7&  3--4&  6  & 9\\
\hline \hline
\end{tabular}

$^a$ The number of sources per bin per optical type: first the number of type 1 sources and second the number of not type 1 sources. 
$^{aa}$ The number of sources per bin included in the lower and upper boundaries are reported. 
\label{xo_table}
\end{table}

\section{Discussion and Conclusion}

In this paper we have selected a sample of 81 high redshift ($z>$3) sources plus 20 candidate high redshift sources in the C-COSMOS survey to 
study the evolution and space density of high redshift AGNs in both the high and low luminosity regime. 
This sample is the largest available sample of  $z>$3 X-ray selected AGNs in a contiguous sky area. 
Using the photometric redshifts and their associated {\it P(z)}, we also were able to compute the 
effective size of the sample (73 sources), by summing the probability of being at z$>$3 of all the sources.

The number counts have been derived in the soft 0.5-2 keV band (observed), while the space density has been computed in the 
hard 2-10 keV band (rest frame), to minimize the bias introduced by the obscuration.
Errors associated with the uncertainties on the sensitivity of the survey and the assumption on the X-ray spectral shape 
are taken into accounts. 

Studying the number counts of high redshift quasars can constrain the evolution of the BH properties. 
Comparing physical models with number counts could be more efficient than comparing directly with the luminosity
function which is a derived quantity that is affected by the selection function.
The observed number counts at $z>$3 and $z>$4 
are better reproduced by models with a prolonged quasar light curve characterized by
a long, post-peak activity phase (dot-dashed line in Fig.~3, Shankar 2010),  
and a higher minimum halo mass hosting quasars with respect to a model with smaller halo mass (e.g., Shen 2009; 
dashed line in Fig.~3).
Alternatively, we found that a model characterized by
lower minimum host halo masses and negligible
post-peak phase can equally well reproduce the number counts (dotted line in Fig.~3). 
A combined analysis with number counts and 
clustering measurements (e.g., Allevato et al. 2011) in this redshift and luminosity range
should help to break these physical model degeneracies.

Taking advantage of the large number of sources and the depth of the survey, we are able to probe 
the space density of X-ray selected quasars up to $z\sim$5 at high luminosity (log~L$_{(2-10keV)}>$44.15~\lum). 
The combination of soft, hard and full band detected sources allows us to take into account 
for the presence of highly obscured sources in the derivation of the space density with limited model assumptions.

The comoving space density is in agreement with the predictions from the LDDE Gilli et al. (2007) model at 
high X-ray luminosity and at all redshifts, confirming the declining space density as observed in 
the optical. Doubling the sample, at least, in the $z>$5 and  $z>$6 bins would strengthen the agreement with 
the Gilli et al. (2007) model, now based on a sample of 5 sources.
The LADE Aird et al. (2010) model over estimates the space density at $z>$4, even when the errors are taken into account.  

The flux limit of the survey, thus the lack of sources beyond $z>$3.5 in the low luminosity regime, does not yet allow 
strong constraints to be put on the density evolution at the faint end (log~L$_{(2-10keV)}<$44.15~\lum), where somewhat different results have been 
recently found in the optical band by Glikman et al. (2010, 2011) and Ikeda et al. (2011), 
with respect to previous studies (e.g., Fontanot et al. 2007).
Despite our sample of 32 source at 3$<z<$3.5 in the low luminosity range, it is not possible to give firm results 
when comparing the data with models. 
To obtain a more conclusive understanding of the quasar evolution in the early universe, in particular 
at $z>$3.5, where the models diverge strongly, larger and complete samples of low-luminosity AGNs are required.

\begin{figure}
\centering
\includegraphics[width=0.49\textwidth]{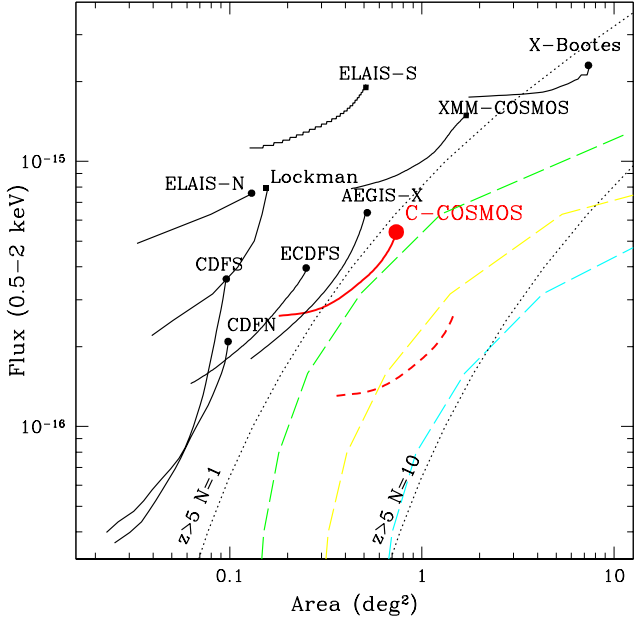}
\caption{ Area-flux curves for Chandra and XMM-Newton contiguous X-ray surveys (black solid lines). Each survey has been plotted using each sensitivity
curve starting from the flux corresponding to the area that is 80\% of the
maximum area for that survey (large points at the top of each curve), to
the flux corresponding to the 20\% of the total area (bottom of each curve). 
For the references of each survey see Fig.~5 of Elvis et al. (2009). The red solid and dashed lines represent the C-COSMOS survey and 
a potential 6 Ms large \chandra\ COSMOS survey (see text), respectively. The dashed lines show the 
area and flux required to observe 15 sources with luminosity in the range 10$^{43}$-10$^{44}$ \lum, 
in the following redshift bins: $z$=3-3.5 green dashed; 
$z$=3.5-4 yellow dashed; $z$=4-4.5 cyan dashed. The dotted lines show the area and flux required to observe N=1, 10 
bright X-ray luminosity ($>$10$^{44}$ \lum) quasars at z$>$5. }
\label{areaflux}
\end{figure}

We can now quantify the best strategy to obtain these larger samples. 
In Figure \ref{areaflux}, the well known ``area-flux'' plot (e.g., Brandt \& Hasinger 2005) 
used to compare different X-ray surveys is reported as in Elvis et al. (2009; i.e. plotting the flux at which each survey 
reaches from 20\% to 80\% of the area surveyed). 

Using the agreement of the predictions between the Gilli et al.
(2007) model with our data, we estimated curves of constant number of sources (N=15) in 3 redshift bins ($z$=3-3.5 green dashed; 
$z$=3.5-4 yellow dashed; $z$=4-4.5 cyan dashed) for the low 
luminosity range 10$^{43}$-10$^{44}$ \lum, to estimate the flux limit and the area needed to 
produce larger sample of faint AGNs at high redshift. 
These curves show that moving the survey sensitivity diagonally in this plot is a better investment, for these purposes, than 
going sidey for either increased depth or area. 
 
Doubling the coverage of the COSMOS area and increasing the depth by a factor 2 (dashed red thick line) 
would substantially increase the $z>$3 low luminosity AGN statistics by a factor 2 in the 10$^{43}$-10$^{44}$ \lum\ range. 
This would be sufficient to resolve the controversial optical results in this luminosity range, based on small samples.  

A ``double-doubled'' survey of this kind would also increase the sample of z$>5$ quasars (dotted black line) 
from 4 in the soft band to 7-8 sources at high luminosity. 
Such a survey, performed with the same observational strategy of C-COSMOS (i.e. homogeneous exposure time and 
same tiling), would take approximately 6 Ms with \chandra, the scale of the X-ray Visionary Projects. 
In this respect, the proposed Wide Field X-ray Telescope 
(WFXT) Medium survey (Rosati et al. 2010) designed to cover a large area (3000 deg$^2$) at the fluxes of the C-COSMOS survey 
would be optimal to collect order of magnitude larger samples of high redshift AGNs compared to X-ray and also optical surveys 
now available (Brusa et al. 2010; Gilli et al. 2010). 

\acknowledgments
The authors thank the referee for the suggestions which improved the paper content.
The authors thank J. Aird and E. Glikman for sharing their luminosity functions, 
G. Risaliti, R. D'Abrusco and A. Goulding for 
useful discussions. This work was supported in part by NASA Chandra grant
number GO7-8136A (ME, FC, HH), the Blancheflor Boncompagni 
Ludovisi foundation (FC) and the Smithsonian Scholarly Studies (FC). 
FS acknowledges the Alexander von Humboldt 
Foundation for support. In Italy this work is
supported by ASI/INAF contracts I/009/10/0 and
I/088/06.  

{\it Facilities:} \facility{Chandra (ACIS)}, \facility{HST (ACS)}, \facility{XMM}.

\end{document}